\documentclass[conference,11pt]{IEEEtran}
\usepackage[utf8]{inputenc}
\usepackage[english]{babel}
\usepackage{acronym}
\usepackage{amsmath,amssymb,amsfonts}
\usepackage{algorithm}
\usepackage{algpseudocode}
\usepackage{csquotes}
\usepackage{graphicx}
\usepackage[hidelinks]{hyperref}
\usepackage{listings}
\usepackage{subcaption}
\usepackage{textcomp}
\usepackage{xcolor}

\lstdefinestyle{lststyle}{
  captionpos=b,
}
\def\BibTeX{{\rm B\kern-.05em{\sc i\kern-.025em b}\kern-.08em
    T\kern-.1667em\lower.7ex\hbox{E}\kern-.125emX}}

\begin{document}

\title{Domain-specific Hardware Acceleration for Model Predictive Path Integral Control}

\author{Erwan Tanguy-Legac, Tommaso Belvedere, Gianluca Corsini, Marco Tognon, Marcello Traiola\\
Univ Rennes, Inria, CNRS, IRISA, UMR6074
}

\maketitle

\begin{abstract}
	Accurately controlling a robotic system in real time is a challenging problem.
	To address this, the robotics community has adopted various algorithms, such as Model Predictive Control (MPC) and Model Predictive Path Integral (MPPI) control.
	The first is difficult to implement on non-linear systems such as unmanned aerial vehicles, whilst the second requires a heavy computational load.
	GPUs have been successfully used to accelerate MPPI implementations; however, their power consumption is often excessive for autonomous or unmanned targets, especially when battery-powered.
	On the other hand, custom designs, often implemented on FPGAs, have been proposed to accelerate robotic algorithms while consuming considerably less energy than their GPU (or CPU) implementation.
	However, no MPPI custom accelerator has been proposed so far. In this work, we present a hardware accelerator for MPPI control and simulate its execution. Results show that the MPPI custom accelerator allows more accurate trajectories than GPU-based MPPI implementations.
\end{abstract}

\section{Introduction}

Modern robotic applications require the ability to perform complex tasks in real-time, with fast, accurate responses to the environment.
Due to the increase in computational power on modern hardware, techniques such as Model Predictive Control (MPC), previously used in heavy industrial plants, have been adopted on various robotic platforms \cite{nguyen2021model, katayama_model_2023}.
MPC works by selecting the best trajectory for the system over a receding time horizon.
This is done by solving an optimization problem that minimizes a cost function subject to the system's constraints.

Another algorithm that has gained some popularity in the robotics community is Model Predictive Path Integral (MPPI) control \cite{williams_model_2017}.
This algorithm can be viewed as a sampling-based version of MPC, where the optimization problem is solved by sampling multiple trajectories over a receding time horizon and computing a weighted average of them.

Although these algorithms have been successfully implemented across various platforms, they are computationally intensive, requiring high-performance hardware.
This is especially true for non-linear systems, such as Unmanned Aerial Vehicles (UAVs), where the underlying optimization problem is non-convex and expensive to solve, making the use of local methods difficult.
Naturally, the computational power required for such calculations must be taken into account, especially for portable and autonomous systems.
As such, the community has begun porting the MPC algorithm to high-performance, specialized hardware such as GPUs \cite{abughalieh_survey_2019} and FPGAs \cite{joos_parallel_2011}, both of which are highly efficient for parallel computation.

FPGAs have been used to implement MPC on embedded systems with a power consumption lower than 1 Watt \cite{lucia_optimized_2018}, and on many other robotic applications \cite{wan_survey_2021, wu_fpga_2021}.
They can offer non-negligible power reduction and overall performance improvements when compared to CPUs and GPUs in various fields, such as vision kernel computation \cite{qasaimeh_comparing_2019}, image processing \cite{siddiqui_fpga_2019}, and network security \cite{hayajneh_wlan_2017}.
However, to the best of our knowledge, no implementation of MPPI on a dedicated hardware accelerator (other than a GPU) has been presented yet, which is what we address in this work.

The rest of this paper is organized as follows: first, section \ref{sec:mppi} gives an overview of the Model Predictive Path Integral (MPPI) control algorithm.
Then, in section \ref{sec:accel}, we present our hardware accelerator, designed to accelerate the MPPI algorithm by leveraging the capabilities of application-specific integrated circuits.
Finally, we present our experimental results and discuss the limitations of our design.

\section{Model Predictive Path Integral Control} \label{sec:mppi}

\paragraph*{Notations} In this paper, $H$ refers to the length of the time horizon, $N$ refers to the number of sampled trajectories, $x_t$ refers to the state of the system at time $t$, and $u_t$ refers to the inputs to be applied to the system at time $t$.
Additionally, we use two distinct cost functions: $\mathcal L$ for the running cost (computed for each intermediate state of a trajectory), and $\phi$ for the final cost (computed for the last state of a trajectory).
Finally, for each $i \in [\![0, N-1]\!]$, a superscript $i$ is used to refer to the $i^{th}$ sampled trajectory.

\vspace{0.5em}

In robotics, algorithms for real-time trajectory optimization and control in robotics -- such as MPC and MPPI -- aim to compute control inputs that move a robot toward a goal while respecting system dynamics, constraints, and optimizing a performance cost (e.g., energy, smoothness, distance to goal, obstacle avoidance).

As previously stated, MPPI is a sampling-based algorithm that can be seen as a variant of MPC.
Where MPC computes the optimal trajectory by solving a constrained optimization problem, MPPI uses a sampling-based approach.
Such an approach is particularly well-suited to nonlinear systems because it offers gradient-free optimization.
As explained in \cite{williams_information_2018}, it also enables the use of discontinuous cost functions and naturally handles the noise present in the system.

To do so, MPPI can be divided into three distinct steps: first, we generate $N$ sequences of $H$ inputs, randomly sampled from a normal distribution centered on the current input.
Then, for each of these sequences, we predict the system's trajectory if the inputs were applied and compute its cost.
Finally, we compute an approximation of the optimal trajectory by taking a weighted average of the inputs, where the weights are given by the cost of the corresponding trajectory.
The first input of this trajectory is then applied to the system's actuators, and the rest of the sequence is discarded so that a new computation can start.

These three steps are briefly detailed in the rest of this section.
The most computationally intensive step is the \textit{rollout}, which simulates the system's dynamics over the entire time horizon.
It is this specific step that we focus on accelerating in our work (see section \ref{sec:accel}).
The first and last steps are still performed by the CPU.

Algorithm \ref{lst:mppi} shows the pseudocode of the MPPI algorithm.
Its three core steps are described below.

\paragraph{Input sampling (line 1)}

The algorithm starts by sampling $N$ sequences of $H$ inputs from a multivariate normal distribution centered on the current input $v$ with standard deviation $\sigma$.

\paragraph{Rollout (lines 2-10)}

Taking the inputs sampled in the previous step and predicting the resulting trajectory along with its cost is referred to as the \textit{rollout}.
This step is naturally parallelizable, as each trajectory is independent of the others.
To do so, the \textit{rollout} relies on a model of our system, which is used to perform a discrete integration of the system's dynamics over time.
Put in other words, at each time step over the time horizon, we compute the next state of the system based on the current state and the input to be applied (line 6).
Calculating the cost of the corresponding trajectory is also done at each of these time steps, to which we assign a cost value that depends on a reference (consisting of a state and an input, see line 7).
To put emphasis on the result of the trajectory, we compute an additional final cost at the end of the \textit{rollout}, to emphasize the final state of the trajectory (line 9).
These costs are then summed, yielding a final cost $J^i$ for each input sequence $u^i$.

\paragraph{Computing the input to be applied (lines 11-17)}

The input to be applied is computed by taking a weighted average of the input sequences used during the \textit{rollout}.
To each input sequence $u^i$, we assign a weight $w^i$ based on the cost $J^i$ associated with the trajectory it produced.

$$u^* = \frac{\sum_{i=0}^{N-1}w^i u_0^i}{\sum_{i=0}^{N-1}w^i}$$

$$ w^i = \exp\left(-h \frac{J^i - J_\top}{J_\bot - J_\top}\right)$$

Where $J_\top = \max\{J^i\}$ and $J_\bot = \min\{J^i\}$, and $h > 0$ is a parameter to be set.

Finally, $u^* = v' / d$ is applied, and the whole process can start again.

\begin{algorithm}
	\caption{MPPI algorithm} \label{lst:mppi}
	\algrenewcommand\algorithmicrequire{\textbf{input:}}
	\begin{algorithmic}[1]
		\Require current state $x_0$, reference $R$, current input $v$
		\State $u \gets$ sample $N \times H$ inputs from $\mathcal{N}(v, \sigma)$
		\For{$i \textnormal{ in } 0 \textnormal{ to } N-1$}
		\State $x_{0}^i \gets x_0$
		\State $c^i \gets 0$
		\For{$t \textnormal{ in } 0 \textnormal{ to } H-1$}
		\State $x_{t+1}^i \gets F(x_t^i, u_t^i)$
		\State $c^i \gets c^i + \mathcal{L}(x_t^i, u_t^i, R_t)$
		\EndFor
		\State $J^i \gets c^i + \phi(x_{H-1}^i, u_{H-1}^i, R_{H-1})$
		\EndFor
		\State $J_\bot, J_\top \gets \min\{J^i\}, \max\{J^i\}$
		\State $v', d \gets 0, 0$
		\For{$i \textnormal{ in } 0 \textnormal{ to } N-1$}
		\State $v' \gets v' + w^i u_0^i$
		\State $d \gets d + w^i$
		\EndFor
		\State \Return $v' / d$
	\end{algorithmic}
\end{algorithm}

\vspace{0.5em}

Given that the most demanding step of the algorithm is the \textit{rollout}, it is the one we find most interesting to accelerate.
As stated in the introduction, this is already commonly done using GPUs.
However, GPUs are very similar to traditional computing architectures, such as CPUs, but are more efficient in parallel tasks.
As such, they face limitations because they cannot leverage the inner structure of the operations.
Whereas GPUs are very good at parallelization (\textit{i.e.} multiple rollout instances can be run in parallel), they cannot leverage the intrinsic pipelining property of MPPI over the time horizon.
However, this can be addressed using custom accelerators, which enable \textit{pipelining} on top of the parallelism already available in GPUs.

\section{A dedicated hardware accelerator for MPPI} \label{sec:accel}

MPPI is well-suited for parallelization, as each rollout is independent of the others.
As such, GPUs are commonly employed to perform most of the computation required by MPPI, due to their excellent performance when processing parallel tasks \cite{turrisi_mppi_2024}.
This can be replicated on a dedicated hardware accelerator by instantiating multiple computation units, each computing a single rollout.

\begin{figure*}[h]
	\includegraphics[width=\textwidth]{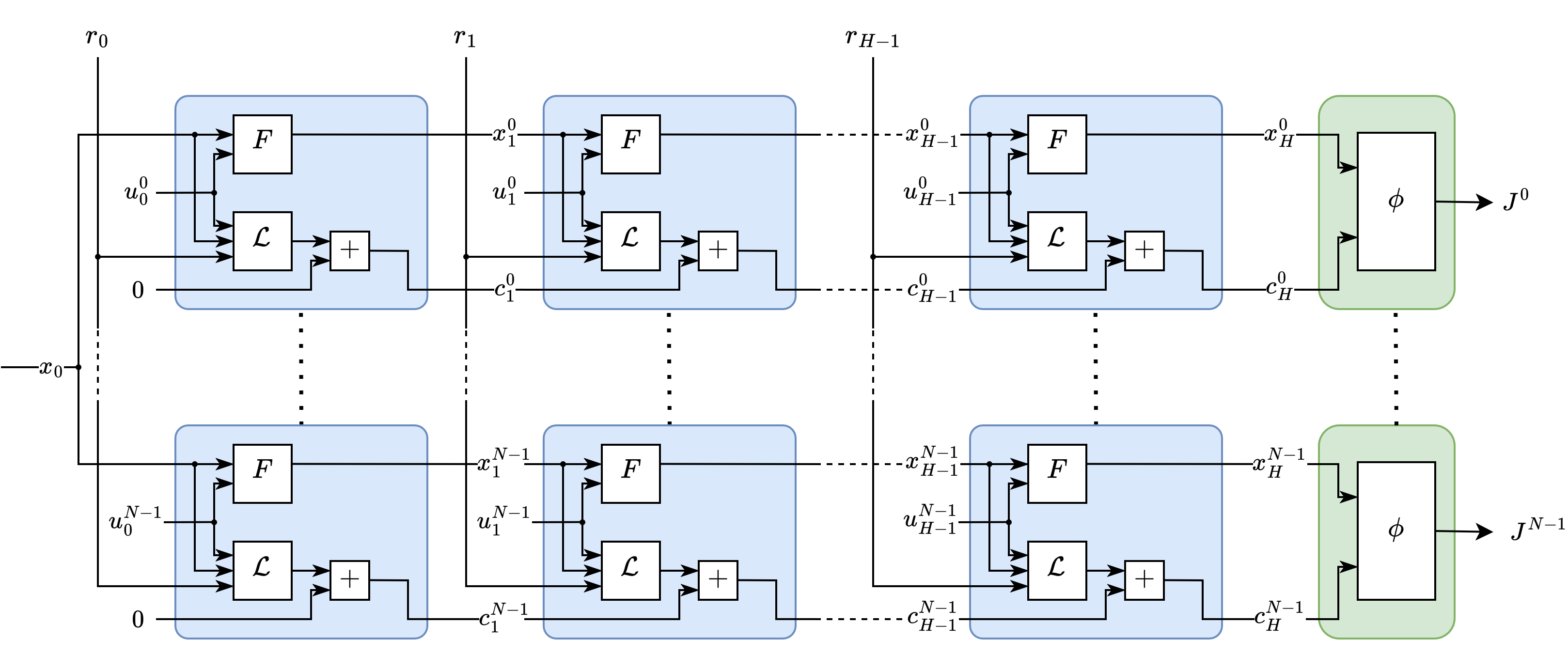}
	\caption{Architecture of the proposed accelerator.}
	\label{fig:architecture}
\end{figure*}

Dedicated hardware accelerators offer more acceleration opportunities than GPUs can.
The next two subsections dive deeper into the algorithm to present the acceleration opportunities that we implemented in our design.
Figure \ref{fig:architecture} shows the final architecture of our accelerator.

\subsection{Pipelining an entire rollout}

The rollout computation is primarily performed in a loop, where each iteration requires only the result of the previous iteration (plus some external data that remains constant throughout execution).
Thanks to this property, we can unroll the loop and implement it using a pipeline.
A pipelined loop does not have to wait for an entire loop execution to finish before starting the next one.
Once the first iteration is completed, the data can be moved to the next stage of the pipeline, and a new input can be accepted in the first stage.

\begin{figure}
	\includegraphics[width=\columnwidth]{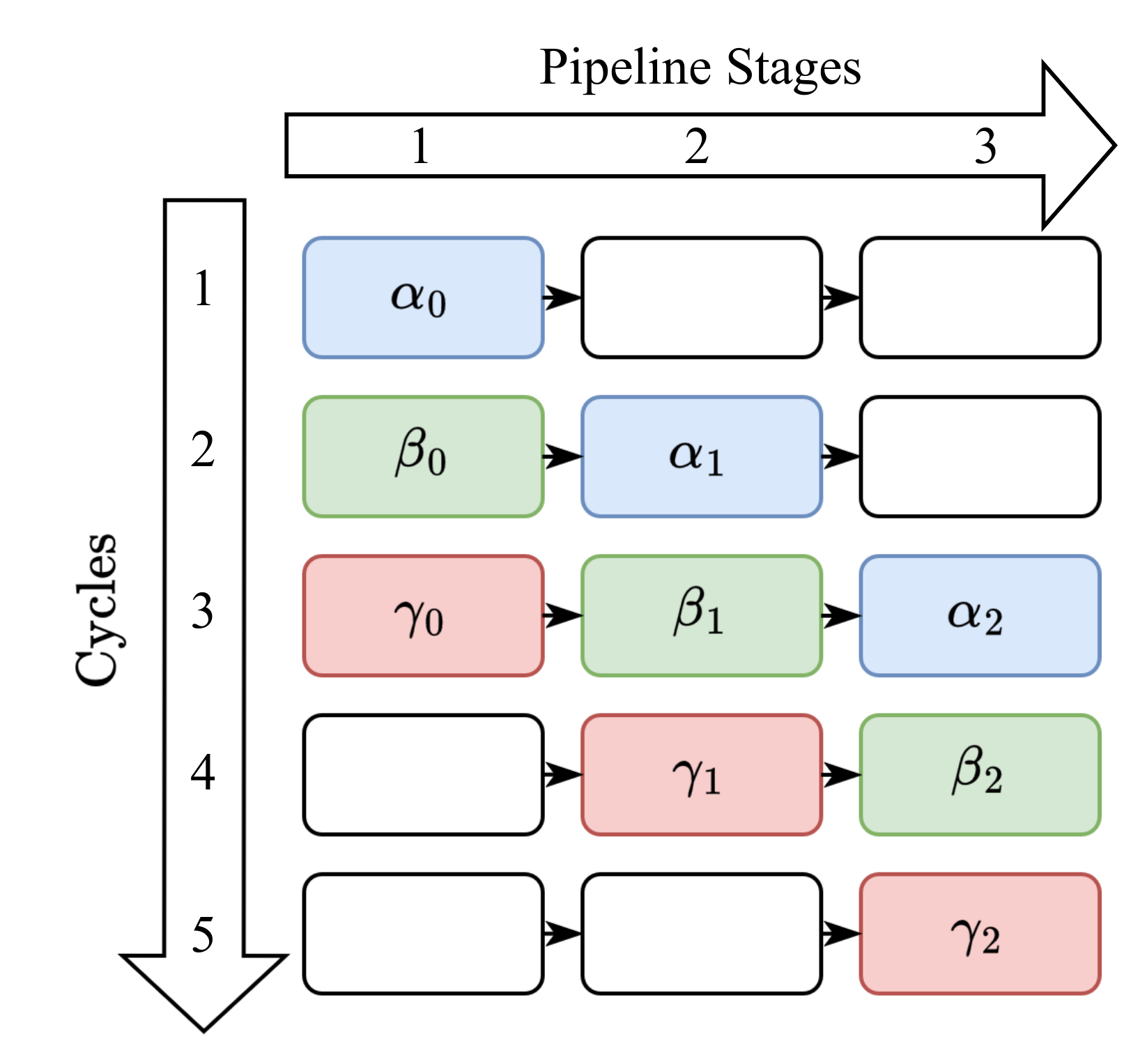}
	\caption{Example of a three stage pipeline applied to the sequence $(\alpha_0, \beta_0, \gamma_0)$}
	\label{fig:pipeline}
\end{figure}

As an example, let us consider a short horizon of three steps ($H=3$) and the rollout of a short input sequence $(\alpha_0, \beta_0, \gamma_0)$. In the conventional (i.e., non-pipelined) way of computing a rollout, each input would go through 3 iterations before obtaining the final rollout result. To compute the entire rollout sequence on the 3-input sequence, 9 iterations are necessary.
To show the advantage of pipelining, Figure \ref{fig:pipeline} shows a simple example of a 3-stage  pipeline, each performing the operation $x_i \mapsto x_{i+1}$.
The first iteration starts by loading $\alpha_0$ in the pipeline's first stage.
At the second iteration, the first stage has completed its computation, the result $\alpha_1$ is processed by the second stage, and the first stage can accept the new input $\beta_0$. By following this approach, after 5 iterations, the whole input sequence has been processed, and the rollout is complete. For a k-input sequence, the number of iterations for the non-pipelined version would be $3k$ while for the pipelined version $k+2$  (the first result appears after 3 cycles, then one result per cycle). In general, for a horizon $H$, the pipeline requires $k + H - 1$ iterations while the non-pipelined version requires $H\cdot k$ iterations. For large input sequences $k$, the speedup of the pipelined version is $\lim_{k   \rightarrow \infty} \frac{H\cdot k}{k+H-1}\approx H$ (i.e., $\frac{3\cdot k}{k+3-1}\approx 3$ for the example in the figure).

We applied this concept to our design by pipelining the rollout loop. The final design uses $N$ pipelines with $H+1$ stages, where the final cost computation is performed on a distinct pipeline stage.

In summary, a pipeline's main advantage is that it allows us to compute the next state of the system while the previous one is still being computed.
This results in a higher throughput, as the system can accept new inputs while the previous ones are still being processed.
However, the total number of loop iterations remains the same, so the delay does not change much when comparing a GPU with our accelerator.

\subsection{Accelerating a single pipeline stage}

Our architecture is now composed of a sequence of pipeline stages, each computing a single rollout iteration.
Additionally, the computation of the error and the next state are independent, and can be parallelized, as shown in Figure \ref{fig:architecture}.

From a system perspective, the reference $R$ and current input $v$ are transferred by the host CPU to the accelerator, and the current state $x_0$ of the system is read from the sensors.
Each pipeline computes the cost of the trajectory $\left[x_0, F(x_0, u_0^i), F(F(x_0, u_0^i), u_1^i), \dots\right]$ and sends it back to the CPU.
The CPU is therefore responsible for computing the final trajectory.

\section{Experimental evaluation}

In this section, we present simulation results obtained with our proposed accelerator architecture, along with synthesis results.

\subsection{Simulation}

\begin{figure}[h]
\centering
	\begin{subfigure}{0.9\columnwidth}
		\includegraphics[width=\columnwidth]{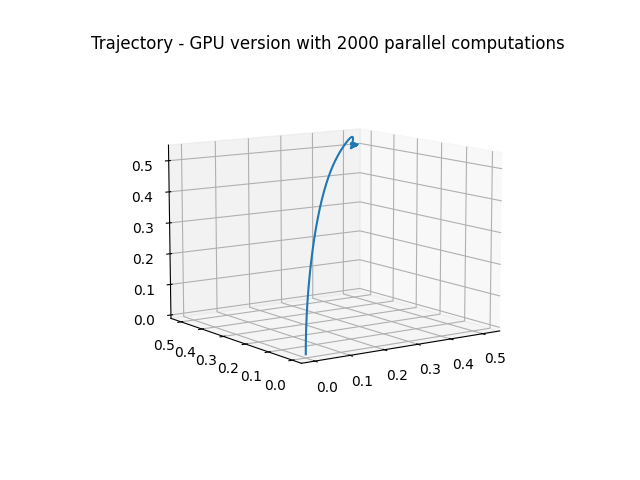}
	\end{subfigure}
	\hfill
	\begin{subfigure}{0.9\columnwidth}
		\includegraphics[width=\columnwidth]{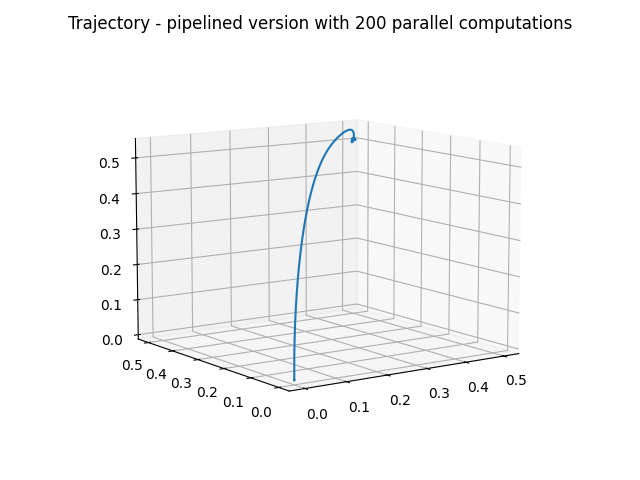}
	\end{subfigure}\caption{
		Simulation without obstacles.
		Position is plotted in 3D space.
		Above, the simulated system contains a GPU executing 2000 parallel rollouts, whilst at the bottom the system is simulated as if it had a FPGA with 200 parallel pipelines.
	}
	\label{fig:simtraj}
\end{figure}

As previously mentioned, the pipelined architecture of our accelerator enables us to increase the system's throughput.
To show the effect of the proposed MPPI custom pipelined acceleration on unmanned aerial vehicles, we simulated the dynamics of a quadrotor when using the proposed MPPI acceleration implementation.
The first simulated task was to reach a target position $(x, y, z) = (0.5, 0.5, 0.5)$.
To run these simulations, we use a model $F$ of the quadrotor which, given a state and a set of inputs, produces a new state.
A state consists of position, rotation, acceleration, and angular velocity, and the inputs are the torques to be applied on each axis, along with a global thrust value.
Each simulation was then executed for 10 seconds, with a receding horizon composed of 25 intervals of 0.02 seconds each (i.e., the rollouts predicted trajectories 0.5 seconds in advance). Figure~\ref{fig:simtraj} shows the trajectory obtained in both cases.

The GPU-based implementation runs with $N = 2000$ parallel rollouts, whilst the hardware-accelerated one is limited to $N = 200$.
This is done to emphasize the fact that hardware space for dedicated accelerators is often limited.
However, because our accelerator uses a pipeline, it achieves much higher throughput.


\begin{figure}[h]
\centering
	\begin{subfigure}{0.49\columnwidth}
		\includegraphics[width=\columnwidth]{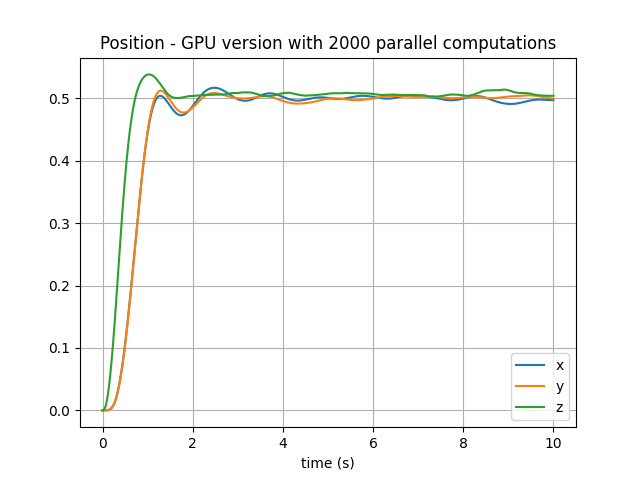}
	\end{subfigure}
	\hfill
	\begin{subfigure}{0.49\columnwidth}
		\includegraphics[width=\columnwidth]{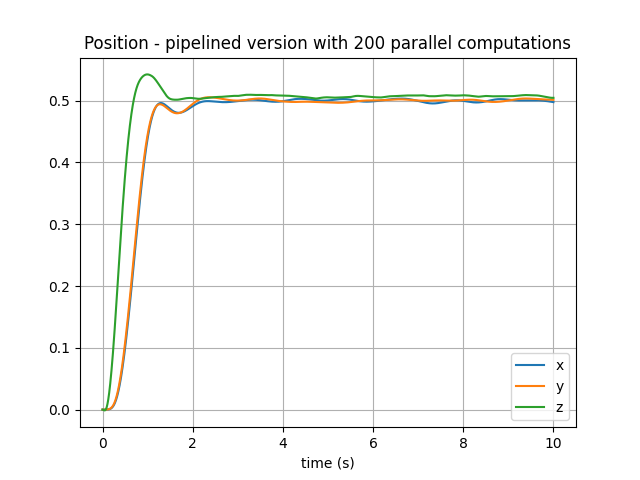}
	\end{subfigure}

	\begin{subfigure}{0.49\columnwidth}
		\includegraphics[width=\columnwidth]{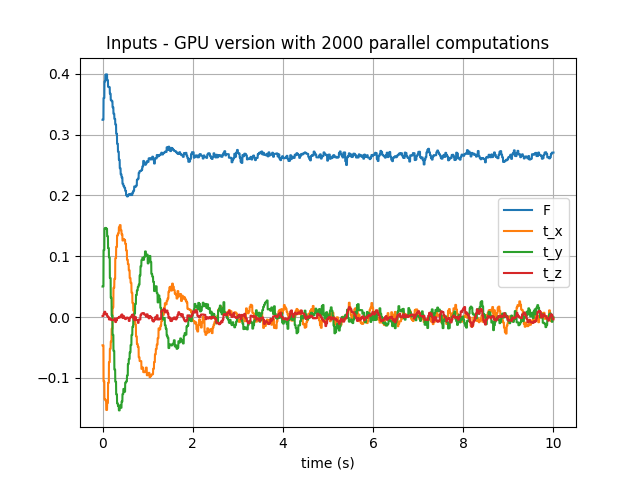}
	\end{subfigure}
	\hfill
	\begin{subfigure}{0.49\columnwidth}
		\includegraphics[width=\columnwidth]{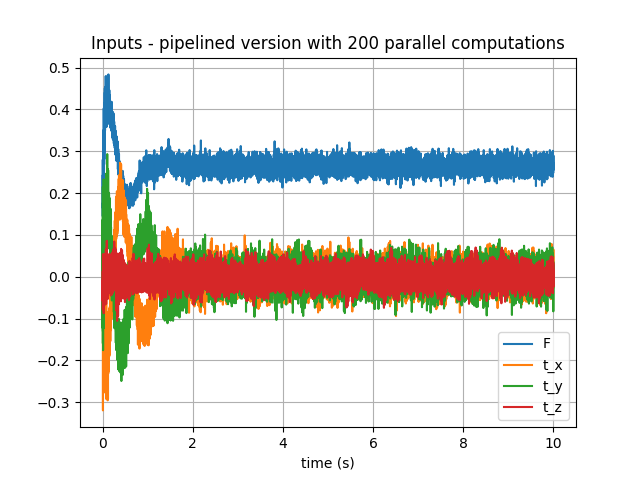}
	\end{subfigure}
	\caption{
		Simulation without an obstacle.
		On the left, the simulated system contains a GPU executing 2000 parallel rollouts, whilst on the right the system is simulated as if it had a FPGA with 200 parallel pipelines.
		The first row contains position plotted against time, and the second row presents the inputs (\textit{i.e.} commands sent to the actuators), also plotted against time.
	}
	\label{fig:sim}
\end{figure}

Figure \ref{fig:sim} presents the results of the simple simulation as described at the start of the section.
It is interesting to note that although the pipelined version involves 10 times fewer parallel executions of the rollout step (and, as a result, 10 times fewer trajectories are sampled), it produces much smoother trajectories and better stabilizes at its target position.
However, it achieves this by having much finer control over the system's actuators, which may not always be possible.
This is especially noticeable on the plots of the input values, which vary much more when using the pipelined version.

\begin{figure}
\centering
	\begin{subfigure}{0.8\columnwidth}
		\includegraphics[width=\columnwidth]{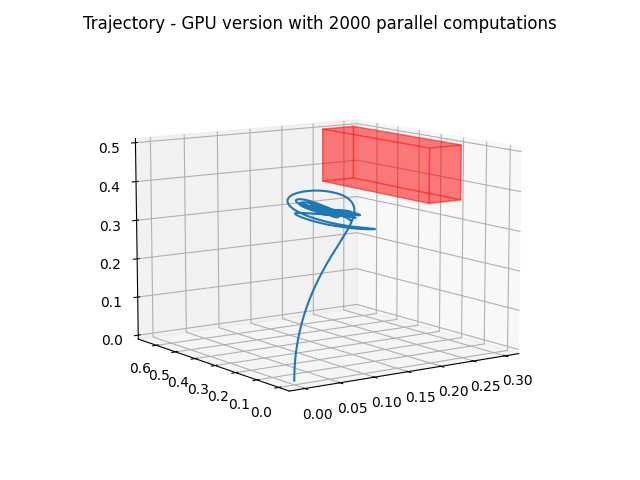}
	\end{subfigure}
	\hfill
	\begin{subfigure}{0.8\columnwidth}
		\includegraphics[width=\columnwidth]{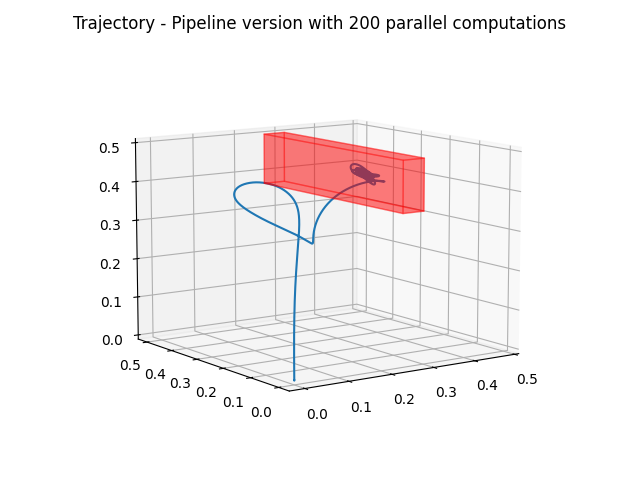}
	\end{subfigure}
	\caption{
		Simulation with an obstacle.
		Position is plotted in 3D space; the red box is an obstacle.
		Above, the simulated system contains a GPU executing 2000 parallel rollouts, whilst at the bottom the system is simulated as if it had a FPGA with 200 parallel pipelines.
	}
	\label{fig:obs}
\end{figure}

The second simulated scenario we ran contains a simple static obstacle, positioned on the trajectory defined in the first simulation.
Figure \ref{fig:obs} shows the results of this simulation: the GPU-based MPPI implementation fails to find a suitable trajectory for the UAV to reach its goal, whereas the proposed MPPI custom acceleration successfully finds one.

Our simulations show that, assuming the computation time of a single rollout iteration to be at most ten times slower on our accelerator than on a GPU (taking data transfer into account), and the ability to synthesise 200 parallel instances of the pipeline, the use of a pipelined architecture can offer significant improvements on the trajectory of our system.
However, in complex dynamic systems such as quadrotors, simulation cannot fully replace real-world tests. Hence, before running such tests, the design must be synthesized for deployment on a real target hardware platform.

\subsection{Design synthesis}

Two options for hardware synthesis exist: reprogrammable circuits such as FPGAs, which enable rapid prototyping and testing of new designs, and Application-Specific Integrated Circuits (ASICs), which are more efficient but require a longer, more expensive design cycle.

\begin{figure}
	\includegraphics[width=\columnwidth]{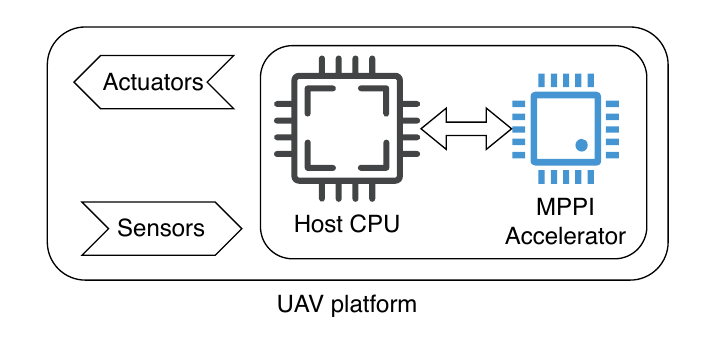}
	\caption{High-level view of the system.}
	\label{fig:computing}
\end{figure}

\begin{table*}[ht]
    \centering
    \begin{tabular}{|c|c|c|c|}
        \hline
            \textbf{Number of pipelines} & \textbf{Number of stages per pipeline} & \textbf{Size of a stage} & \textbf{Utilization (\% of LUTs)} \\ \hline
            $1$ & $25$ & $1$ & $12.58 \%$ \\ \hline
            $5$ & $25$ & $1$ & $83.28 \%$ \\ \hline
            $2$ & $5$ & $5$ & $66.30 \%$ \\ \hline
            $10$ & $1$ & $25$ & $69.22 \%$ \\ \hline
            $14$ & $1$ & $25$ & $96.56 \%$ \\ \hline
    \end{tabular}
    \caption{Utilization results for an \textit{Alveo U55C datacenter FPGA} using \textit{Vivado 2023.1}'s default synthesis settings.}
    \label{tab:synth}
\end{table*}

Figure \ref{fig:computing} shows a high-level view of the proposed system, which is composed of a host (a CPU or microcontroller) that offloads the computation of the rollout part of the MPPI algorithm to a hardware accelerator 
and interacts with the environment using sensors and actuators.

Our implementation was written in \textit{VHDL} and \textit{Verilog}.
Synthesis was performed on an \textit{Alveo U55C datacenter FPGA} using \textit{Vivado 2023.1}.
It is worth noting that our design uses 32-bit floating-point representation for values, although the \textit{U55C} does not contain an FPU.
This can explain the high LUT usage presented in Table \ref{tab:synth}.

Due to the high resource usage of a single pipeline instance ($12.58 \%$), we decided to introduce a new tradeoff: instead of having each pipeline stage compute a single rollout iteration, we can make each of them compute multiple rollout iterations. 
In Table \ref{tab:synth}, we refer to the number of iterations performed by each pipeline stage as the size of the stage.
This way, we can reduce the total number of pipeline stages while keeping the same time horizon for trajectory generation.
This modification introduces only minor overhead at each of these stages.

However, this trade-off may not be necessary when targeting ASICs, as they do not have a fixed number of LUTs.
ASICs would enable larger designs, albeit at the cost of losing the reprogrammable feature, making them suitable for deployment at the end of the design development process.
Moreover, the specialized nature of ASICs enables even more efficient design, further reducing power consumption~\cite{amara_fpga_2006}.



\section{Conclusion}

In conclusion, this paper demonstrates the possibility of accelerating the MPPI algorithm using a dedicated hardware accelerator.
It shows that such an accelerator can discover smoother trajectories in a system.
The reputation of dedicated hardware accelerators also suggests that their use should reduce the algorithm's power consumption.

Future work could propose better hardware implementations, further reducing resource usage and power consumption.
This work also opens the possibility of modifying the MPPI algorithm to better accommodate the acceleration opportunities offered by the hardware.
Finally, it would also be interesting to extend the accelerator to the entire algorithm, and not just the \textit{rollout} step.

\bibliographystyle{IEEEtran}
\bibliography{biblio.bib}

@inproceedings{nguyen2021model,
  title = {Model predictive control for micro aerial vehicles: A survey},
  author = {Nguyen, Huan and Kamel, Mina and Alexis, Kostas and Siegwart, Roland
            },
  booktitle = {2021 European Control Conference (ECC)},
  pages = {1556--1563},
  year = {2021},
  organization = {IEEE},
  doi = {10.23919/ECC54610.2021.9654841},
}

@article{siddiqui_fpga_2019,
  AUTHOR = {Siddiqui, Fahad and Amiri, Sam and Minhas, Umar Ibrahim and Deng,
            Tiantai and Woods, Roger and Rafferty, Karen and Crookes, Daniel},
  TITLE = {FPGA-Based Processor Acceleration for Image Processing Applications},
  JOURNAL = {Journal of Imaging},
  VOLUME = {5},
  YEAR = {2019},
  NUMBER = {1},
  ARTICLE-NUMBER = {16},
  DOI = {10.3390/jimaging5010016},
}

@article{hayajneh_wlan_2017,
  author = {Hayajneh, Thaier and Ullah, Sana and Mohd, Bassam J. and Balagani,
            Kiran S.},
  journal = {IEEE Systems Journal},
  title = {An Enhanced WLAN Security System With FPGA Implementation for
           Multimedia Applications},
  year = {2017},
  volume = {11},
  number = {4},
  pages = {2536-2545},
  doi = {10.1109/JSYST.2015.2424702},
}

@article{katayama_model_2023,
  author = {Sotaro Katayama and Masaki Murooka and Yuichi Tazaki and},
  title = {Model predictive control of legged and humanoid robots: models and
           algorithms},
  journal = {Advanced Robotics},
  volume = {37},
  number = {5},
  pages = {298--315},
  year = {2023},
  doi = {10.1080/01691864.2023.2168134},
}

@article{williams_model_2017,
  title = {Model Predictive Path Integral Control: From Theory to Parallel
           Computation},
  volume = {40},
  doi = {10.2514/1.G001921},
  pages = {344--357},
  number = {2},
  journaltitle = {Journal of Guidance, Control, and Dynamics},
  author = {Williams, Grady and Aldrich, Andrew and Theodorou, Evangelos A.},
  date = {2017-02},
}

@article{abughalieh_survey_2019,
  title = {A Survey of Parallel Implementations for Model Predictive Control},
  volume = {7},
  doi = {10.1109/ACCESS.2019.2904240},
  pages = {34348--34360},
  journaltitle = {{IEEE} Access},
  author = {Abughalieh, Karam M. and Alawneh, Shadi G.},
  date = {2019},
}

@article{lucia_optimized_2018,
  author = {Lucia, Sergio and Navarro, Denis and Lucia, Oscar and Zometa, Pablo
            and Findeisen, Rolf},
  journal = {IEEE Transactions on Industrial Informatics},
  title = {Optimized FPGA Implementation of Model Predictive Control for
           Embedded Systems Using High-Level Synthesis Tool},
  year = {2018},
  volume = {14},
  number = {1},
  pages = {137-145},
  doi = {10.1109/TII.2017.2719940},
}

@inproceedings{qasaimeh_comparing_2019,
  author = {Qasaimeh, Murad and Denolf, Kristof and Lo, Jack and Vissers, Kees
            and Zambreno, Joseph and Jones, Phillip H.},
  booktitle = {2019 IEEE International Conference on Embedded Software and
               Systems (ICESS)},
  title = {Comparing Energy Efficiency of CPU, GPU and FPGA Implementations for
           Vision Kernels},
  year = {2019},
  volume = {},
  number = {},
  pages = {1-8},
  doi = {10.1109/ICESS.2019.8782524},
}

@article{wu_fpga_2021,
  author = {Wu, Yakun and Luo, Li and Yin, Shujuan and Yu, Mengqi and Qiao, Fei
            and Huang, Hongzhi and Shi, Xuesong and Wei, Qi and Liu, Xinjun},
  title = {An FPGA Based Energy Efficient DS-SLAM Accelerator for Mobile Robots
           in Dynamic Environment},
  journal = {Applied Sciences},
  volume = {11},
  year = {2021},
  number = {4},
  doi = {10.3390/app11041828},
}

@article{williams_information_2018,
  title = {Information-Theoretic Model Predictive Control: Theory and
           Applications to Autonomous Driving},
  volume = {34},
  doi = {10.1109/TRO.2018.2865891},
  pages = {1603--1622},
  number = {6},
  journaltitle = {{IEEE} Transactions on Robotics},
  author = {Williams, Grady and Drews, Paul and Goldfain, Brian and Rehg, James
            M. and Theodorou, Evangelos A.},
  date = {2018-12},
}

@article{wan_survey_2021,
  author = {Wan, Zishen and Yu, Bo and Li, Thomas Yuang and Tang, Jie and Zhu,
            Yuhao and Wang, Yu and Raychowdhury, Arijit and Liu, Shaoshan},
  journal = {IEEE Circuits and Systems Magazine},
  title = {A Survey of FPGA-Based Robotic Computing},
  year = {2021},
  volume = {21},
  number = {2},
  pages = {48-74},
  doi = {10.1109/MCAS.2021.3071609},
}

@incollection{joos_parallel_2011,
  title = {Parallel Implementation of Constrained Nonlinear Model Predictive
           Controller for an {FPGA}-Based Onboard Flight Computer},
  pages = {273--286},
  booktitle = {Advances in Aerospace Guidance, Navigation and Control},
  author = {Joos, Alexander and Fichter, Walter},
  date = {2011},
  doi = {10.1007/978-3-642-19817-5_22},
}

@inproceedings{turrisi_mppi_2024,
  author = {Turrisi, Giulio and Modugno, Valerio and Amatucci, Lorenzo and
            Kanoulas, Dimitrios and Semini, Claudio},
  booktitle = {2024 IEEE/RSJ International Conference on Intelligent Robots and
               Systems (IROS)},
  title = {On the Benefits of GPU Sample-Based Stochastic Predictive Controllers
           for Legged Locomotion},
  year = {2024},
  pages = {13757-13764},
  doi = {10.1109/IROS58592.2024.10801698},
}

@article{amara_fpga_2006,
  title = {FPGA vs. ASIC for low power applications},
  journal = {Microelectronics Journal},
  volume = {37},
  number = {8},
  pages = {669-677},
  year = {2006},
  doi = {https://doi.org/10.1016/j.mejo.2005.11.003},
  author = {Amara Amara and Frédéric Amiel and Thomas Ea},
}

\end{document}